\documentclass[aps,prd,preprint,nofootinbib,11pt]{revtex4}
\usepackage{amsfonts}
\usepackage{mathrsfs}
\usepackage{graphicx}
\usepackage{amsmath}
\usepackage{amssymb}
\usepackage{multirow}
\usepackage{subfigure}
\usepackage{epsfig}
\usepackage{graphicx}
\usepackage{color}
\newcommand{\bqa}{\begin{eqnarray}}
\newcommand{\eqa}{\end{eqnarray}}
\newcommand{\beq}{\begin{equation}}
\newcommand{\eeq}{\end{equation}}
\allowdisplaybreaks[1]
\graphicspath{{fig/}{dia/}} \DeclareGraphicsExtensions{.eps}
\begin{document}

\title{Mass Predictions of Vector ($1^{--}$) Double-gluon Heavy Quarkonium Hybrids from QCD Sum Rules\\[0.7cm]}
\author{Chun-Meng Tang$^{1}$,Yi-Cheng Zhao$^{1}$,  and Liang Tang$^{1, 2}$\footnote{tangl@hebtu.edu.cn}}

\affiliation{$^1$ College of Physics, Hebei Normal University, Shijiazhuang 050024, China\\
$^2$ Hebei Key Laboratory of Photophysics Research and Application, Hebei Normal University, Shijiazhuang 050024, China
}

%\author{~\\~\\}

%\affiliation{}

\begin{abstract}
\vspace{0.3cm}
In this work, we study the $1^{--}$ double-gluon charmonium ($\bar{c}ggc$) and bottomonium ($\bar{b} gg b$) hybrids in terms of QCD sum rules. We find that the mass of $\bar{c}ggc$ hybrid lies in $M_{H_{c}}$ = $5.33 \sim 5.90$ GeV, while in the bottom sector the mass of $\bar{b}ggb$ hybrid may be situated in $M_{H_b} = 11.20 \sim 11.68$ GeV. The contributions up to dimension eight at leading order of $\alpha_s$ (LO) in the operator product expansion are taken into account in the calculation. The double-gluon charmonium hybrid meson predicted in this work can decay into a pair of charmed mesons or a pair of charmed mesons together with a light meson. Especially, we propose to search for $\bar{c}ggc$ hybrid with $I^G(J^{PC})= 0^-(1^{--})$ in their decay channels $D \bar{D}/D^* \bar{D}/D^{*} \bar{D}^{*}$ with P wave and $D^* \bar{D}^{*} \pi/D^* \bar{D}^*\eta/D \bar{D}\rho/D \bar{D}\omega$ with S wave, which may be accessible in Belle II, PANDA, Super-B, GlueX, and LHCb experiments.
\end{abstract}
 \maketitle
\newpage

\section{Introduction}\label{Formalism}
Various hadronic structures beyond the normal mesons and baryons are allowed in the framework of quantum chromodynamics (QCD)~\cite{Gross:1973id, Politzer:1973fx, Wilson:1974sk} and quark model~\cite{GellMann:1964nj, Zweig}, such as multiquark states, glueballs, and hybrids, which are nominated as exotic states. A multiquark state is composed of more than three quarks and anti-quarks; a glueball is composed of entirely gluons; a hybrid state contains valence gluon(s), besides valence quarks. Exploring the existence and properties of such exotic states is one of the most intriguing research topics of hadronic physics. In the past two decades, with the development of technology, the research on multiquark states has made tremendous developments, such as the observations of the charmonium-like/bottomonium-like XYZ states~\cite{Belle:2003nnu, BaBar:2005hhc, Belle:2011aa, BESIII:2013ris, Belle:2013yex} and the hidden-charm pentaquarks ($P_c$ states)~\cite{LHCb:2015yax, LHCb:2019kea}( see~\cite{Chen:2016qju, Guo:2017jvc, Olsen:2017bmm, Brambilla:2019esw} for recent reviews), and new ones tend to appear more frequently.

These successes of the XYZ and $P_c$ states have inspired the search for hybrids within the charmonium and bottomonium sectors~\cite{Olsen:2009gi, Olsen:2009ys, Godfrey:2008nc, Close:2007ny}. It is one of the most important design goals to detect the existence of hybrids in many experimental facilities such as BESIII, GlueX, PANDA and LHCb. However, although experimentally the existence of hybrid states has not yet been proved, there are indeed some good candidates observed both recently and in the past. Very recently, the BESIII collaboration reported the first observation of an abnormal state with the exotic quantum number $I^G J^{PC} = 0^+ 1^{-+}$ in the $\eta \eta'$ invariant mass spectrum with a statistical significance larger than 19$\sigma$, named as $\eta_1(1855)$~\cite{BESIII:2022riz, BESIII:2022qzu}. In the past, there were three candidates observed in experiments with the exotic quantum number $I^G J^{PC} = 1^- 1^{-+}$, i.e., the $\pi_1(1400)$~\cite{IHEP-Brussels-LosAlamos-AnnecyLAPP:1988iqi}, $\pi_1(1600)$~\cite{E852:2001ikk, COMPASS:2009xrl}, and $\pi_1(2015)$~\cite{E852:2004gpn}. It is worthy to note that the $\eta_1(1855)$ is the isoscalar partner of the isovector states $\pi_1(1400)$ and $\pi_1(1600)$.

In the past several decades, there accumulated a lot of theoretical studies on hybrids based on various phenomenological models. For example, they have been studied through the MIT bag model~\cite{Barnes:1977hg, Hasenfratz:1980jv, Chanowitz:1982qj}, flux-tube model~\cite{Close:1994hc, Barnes:1995hc, Page:1998gz}, constituent gluon model~\cite{Horn:1977rq, Szczepaniak:2001rg, Guo:2007sm}, AdS/QCD model~\cite{Andreev:2012hw, Bellantuono:2014lra}, lattice QCD~\cite{Michael:1985ne, Perantonis:1990dy, Lacock:1996ny, MILC:1997usn, Juge:1999ie, Juge:2002br, Liu:2005rc, Luo:2005zg, Dudek:2009qf, Liu:2011rn, HadronSpectrum:2012gic, Dudek:2013yja}, and QCD sum rules~\cite{Balitsky:1982ps, Govaerts:1983ka, Govaerts:1984hc, Govaerts:1985fx, Govaerts:1986pp, Kisslinger:1995yw, Zhu:1998ki, Jin:2002rw, Narison:2009vj, Huang:2010dc, Chen:2010ic, Qiao:2010zh, Harnett:2012gs, Berg:2012gd, Chen:2013zia, Chen:2013pya, Kleiv:2014kua, Palameta:2017ols, Palameta:2018yce, Li:2021fwk, Chen:2022qpd}. Among those techniques, QCD sum rules innovated by Shifman, Vainshtein, and Zakharov (SVZ)~\cite{Shifman:1978bx, Shifman:1978by, Reinders:1984sr, Narison:1989aq, P.Col} turns out to be a remarkably successful and powerful technique for the computation of hadronic properties~\cite{Wang:2017qvg, Wang:2019tlw, Chen:2019bip, Tang:2019nwv, Xu:2020evn, Zhang:2020xtb, Albuquerque:2021tqd}. It is a QCD based theoretical framework that incorporates non-perturbative effects universally order by order using the operator product expansion (OPE). In this approach, to establish the sum rules, the first step is to construct the proper interpolating current corresponding to the hadron of interest, which possesses the foremost information about the concerned hadron, such as the quantum number, the constituent quarks and gluons. By using the current, one can then construct the two-point correlation function, which can be investigated at both quark-gluon and hadron levels, usually called the QCD and the phenomenological representations, respectively. After performing the Borel transformation on both representations, we can formally establish the QCD sum rules, from which we can extract the mass of the concerned hadron.

Heavy quarkonium hybrids with one valence gluon ($\bar{Q} g Q$) were originally studied in Refs.~\cite{Govaerts:1984hc, Govaerts:1985fx, Govaerts:1986pp} by Govaerts \emph{et al.}, where they analyzed the masses for various $J^{PC}$ by considering the perturbative and dimension four gluon condensate contributions. Including the tri-gluon condensate contributions to the two-point correlation function, Qiao \emph{et al.} revisited the vector ($1^{--}$) heavy quarkonium hybrids~\cite{Qiao:2010zh}, and found that the tri-gluon condensate contributions can stabilize the hybrid sum rules and allow reliable mass predictions. Then, Chen \emph{et al.} analyzed the heavy quarkonium hybrids with various $J^{PC}$ quantum numbers to include QCD condensates up to dimension six, and drawn similar conclusions~\cite{Chen:2013zia, Chen:2013pya}. Recently, the study of heavy quarkonium hybrid has been extended to calculate the mixing effects between the pure quarkonium hybrids and the quarkonium mesons~\cite{Kleiv:2014kua,Palameta:2017ols, Palameta:2018yce}.

Recently, Chen \emph{et al.} studied a new hadron configuration: the double-gluon hybrid state, which consists of one light quark and one light antiquark together with two valence gluons~\cite{Chen:2021smz}. In this paper, we will study the double-gluon heavy quarkonium hybrid, that is, a pair of heavy quarks and two valence gluons ($\bar{Q} g g Q$). Since a series of newly observed `exotic' states in the charmonium energy region are $J^{PC} = 1^{--}$ hadrons ($Y$ states), it is reasonable to believe that there exist heavier $Y$ states, which may be composed of one charm quark and one anti-charm quark together with two gluons. In this work, we firstly construct four vector ($J^{PC} = 1^{--}$) double-gluon heavy quarkonium hybrid currents. Then, we apply QCD sum rules method to evaluating their masses. Our predictions can be used to analyze the experimental data in the near future.

The rest of the paper is arranged as follows. After the introduction, in Sec. II we derive the formulas of the correlation functions $\Pi_{\mu\nu}(q)$ in terms of the QCD sum rules with the interpolating currents for $J^{PC}=1^{--}$.  The numerical analyses and results are given in Sec. III. Sec. IV is devoted to the decay analyses of the predicted double-gluon charmonium hybrids. The last part is left for conclusions and discussion of the results.

\section{Formalism}\label{Formalism}

In the framework of QCD sum rules, the starting point is to construct the correlation function, \emph{i.e.},
\begin{eqnarray}
\Pi_{\mu\nu}(q)&=&i\int d^{4}xe^{iq\cdot x}\langle 0|T\{j_{\mu}(x),j_{\nu}^{\dagger}(0)\}|0 \rangle,
\end{eqnarray}
where the interpolating current $j_\mu$ for the double-gluon heavy quarkonium hybrids with the quantum number $J^{PC}=1^{--}$ are chosen to be
\begin{eqnarray}
j_{\mu}^{A,1^{--}}(x) &=& g_{s}^{2}f^{abc} G_{\mu\nu}^{a}(x)G^{b\nu\rho}(x)\left[\bar{Q}_{i}(x) (T^{c})_{ij} \gamma_{\rho} Q_{j}(x)\right], \label{currentA} \\
j_{\mu}^{B,1^{--}}(x) &=& g_{s}^{2} f^{abc} \tilde{G}_{\mu\nu}^{a}(x)\tilde{G}^{b\nu\rho}(x)\left[\bar{Q}_{i}(x) (T^{c})_{ij} \gamma_{\rho} Q_{j}(x)\right], \label{currentB} \\
j_{\mu}^{C,1^{--}}(x) &=& g_{s}^{2} f^{abc} G_{\mu\nu}^{a}(x)\tilde{G}^{b\nu\rho}(x)\left[\bar{Q}_{i}(x) (T^{c})_{ij} \gamma_{\rho}\gamma_{5} Q_{j}(x)\right], \label{currentC} \\
j_{\mu}^{D,1^{--}}(x) &=& g_{s}^{2} f^{abc} \tilde{G}_{\mu\nu}^{a}(x)G^{b\nu\rho}(x) \left[\bar{Q}_{i}(x) (T^{c})_{ij} \gamma_{\rho}\gamma_{5} Q_{j}(x)\right],\label{currentD}
\end{eqnarray}
where $g_{s}$ is the strong coupling constant, $i/j=1, 2, 3$ and $a/b/c= 1,2, \cdots, 8$ are color indices, $f^{a b c}$  is the totally antisymmetric $SU(3)$ structure constants, $T^c= \lambda^c/2$ where $\lambda^c$ is the Gell-Mann matrix, $\tilde{G}_{\mu\nu}^{a}(x) = \epsilon_{\mu\nu\alpha\beta} G^{a, \alpha \beta}(x)/ 2$ is the dual field strength of $G_{\mu\nu}^{a}(x)$, and $Q$ represents the heavy-quark $c$ or $b$. Here, the superscripts A to D indicate four different hybrid currents that will be analyzed in our paper.

Generally, the two-point function $\Pi_{\mu\nu}(q)$ may contain two distinct parts, the vector part $\Pi_{V}(q^{2})$ and the scalar part $\Pi_{S}(q^{2})$ which represent the contributions of the correlation function to the vector channel $J^{PC} = 1^{--}$ and scalar channel $J^{PC}= 0^{+-}$, respectively. It can be explicitly expressed as:
\begin{eqnarray}
\Pi_{\mu\nu}(q) &=& (-g_{\mu\nu} + \frac{q_{\mu}q_{\nu}}{q^{2}})\Pi_{V}(q^{2}) + \frac{q_{\mu}q_{\nu}}{q^{2}}\Pi_{S}(q^{2}).
\end{eqnarray}

Since our aim of this work is to study the mass of the vector ($1^{--}$) heavy hybrid, we only analyze the vector part $\Pi_V(q^2)$, which is written as $\Pi(q^2)$ in the following for brevity. The correlation function $\Pi(q^2)$ can be investigated at both quark-gluon and hadron levels, usually called the QCD and the phenomenological representations, respectively. Note that the QCD representation needs analytical calculations, whereas, the mass and coupling constant of the concerned hadron are introduced in the phenomenological representation. In QCD sum rules, the fundamental assumption is the principle of quark-hadron duality, which builds a bridge between the QCD representation and the phenomenological representation, that is:
\begin{eqnarray}
  \Pi^{\text{QCD}}(q^2) = \int_{s_<}^{\infty} d s \frac{\rho^{\text{phen}}(s)}{s-q^2},
\end{eqnarray}
where $\rho^{\text{phen}}(s)$ represents the spectral function on the phenomenological side of QCD sum rules, and the integration starts from the physical threshold. The spectral function, $\rho^{\text{phen}}(s)$, is usually described using some model corresponding to an appropriate resonance shape. In this work, we use the ``one resonance + continuum" approximation for the quark-hadron duality.

At the quark-gluon level, the correlation function can be calculated with the operator product expansion (OPE). As explained in Ref.~\cite{Reinders:1984sr}, it is convenient to introduce the definition of the full propagator of QCD in order to include the non-perturbative effects from QCD vacuum. In our calculation, since we only take into account the contributions up to dimension eight at leading order of $\alpha_s$ in the OPE, it is enough to retain the heavy-quark ($Q=c$ or $b$) full propagator $S^Q_{ij}(p)$ up to single-gluon emission term in momentum space~\cite{Reinders:1984sr}, which is
\begin{eqnarray}
S^Q_{j k}(p) = \frac{i \delta_{j k}(p\!\!\!\slash + m_Q)}{p^2 - m_Q^2} - \frac{i}{4} g_s \frac{t^a_{j k} G^a_{\alpha\beta}(0) }{(p^2 - m_Q^2)^2} [\sigma^{\alpha \beta}
(p\!\!\!\slash + m_Q)
+ (p\!\!\!\slash + m_Q) \sigma^{\alpha \beta}] \; ,\label{prop-heavy}
\end{eqnarray}
where the first term is the perturbative quark propagator, and the second term represents the contribution of the single-gluon emission which forms the gluon condensates $\langle g_s^2 G^2 \rangle$, $\langle g_s^3 G^3 \rangle$, and $\langle g_s^2 G^2 \rangle^2$ together with relevant gluon emission terms from other quark/gluon propagators.

Moreover, the perturbative gluon propagator employed in our analytical calculation is considered in coordinate space, which can be expressed as~\cite{Govaerts:1984hc}:
\begin{eqnarray}
 S_{\mu\nu,\rho\sigma}^{ab}(x)&=& \frac{\delta^{ab}}{2\pi^{2}} \times\frac{1}{x^{6}}\big\{(g_{\mu\rho}x^{2} - 4x_{\mu}x_{\rho})g_{\nu\sigma} - (g_{\mu\sigma}x^{2} - 4x_{\mu}x_{\sigma})g_{\rho\nu}\nonumber\\
 &-& (g_{\rho\nu}x^{2} - 4x_{\rho}x_{\nu})g_{\mu\sigma} + (g_{\nu\sigma}x^{2} - 4x_{\nu}x_{\sigma})g_{\rho\mu}\big\}.\label{pert-gluon}
\end{eqnarray}
Because we work at leading order of $\alpha_s$ and consider condensates up to dimension eight, we also need the gluon propagator associated with single-gluon emission. For simplicity, we shall use it in momentum space, which is derived by ourselves followed Refs.~\cite{Reinders:1984sr, Albuquerque:2012jbz} and has the following expression:
\begin{eqnarray}
S_{\mu\nu, \rho\sigma}^{G, a b}(p)&=& -\frac{i}{2}g_{s} f^{abc_{1}} G^{c_1, \alpha \mu_{1}}(0) \frac{1}{p^{3}} \big\{p^{2}\left[p_{\mu}p_{\rho}(-g_{\alpha\nu}) g_{\mu_{1}\sigma} + p_{\mu}p_{\rho}g_{\alpha\mu_{1}}g_{\nu\sigma} - p_{\nu}p_{\rho}g_{\alpha\mu_{1}}g_{\mu\sigma}\right. \nonumber\\
&+& p_{\nu}p_{\rho} g_{\alpha\mu}g_{\mu_{1}\sigma}
 + g_{\alpha\sigma}\left(p_{\nu}(2p_{\mu_{1}}g_{\mu\rho} - p_{\rho}g_{\mu\mu_{1}}) + p_{\mu} (p_{\rho}g_{\mu_{1}\nu}
 - 2p_{\mu_{1}}g_{\nu\rho})\right)\nonumber\\
  &+& p_{\mu}p_{\sigma}g_{\alpha\nu}g_{\mu_{1}\rho}
  -p_{\mu}p_{\sigma} g_{\alpha\mu_{1}}g_{\nu\rho}  + p_{\nu}p_{\sigma}g_{\alpha\mu_{1}}g_{\mu\rho} - p_{\nu}p_{\sigma}g_{\alpha\mu}g_{\mu_{1}\rho} \nonumber\\
  &+& \left.g_{\alpha\rho} \left(p_{\nu}(p_{\sigma}g_{\mu\mu_{1}} - 2p_{\mu_{1}}g_{\mu\sigma}) + p_{\mu}(2p_{\mu_{1}}g_{\nu\sigma} - p_{\sigma}g_{\mu_{1}\nu})\right) \right]\nonumber\\
  &-& 4p_{\alpha}p_{\mu_{1}}\left( p_{\nu}(p_{\sigma}g_{\mu\rho} - p_{\rho}g_{\mu\sigma}) +p_{\mu}(p_{\rho}g_{\nu\sigma} - p_{\sigma}g_{\nu\rho}) \right)   \big\}.\label{gluon-propagator-single-gluon}
\end{eqnarray}
We refer to Refs.~\cite{Reinders:1984sr,Albuquerque:2012jbz} for the necessary formulas using in the derivation of Eq.\eqref{gluon-propagator-single-gluon}.

\begin{figure}[h]
  \centering
  \includegraphics[width=12cm]{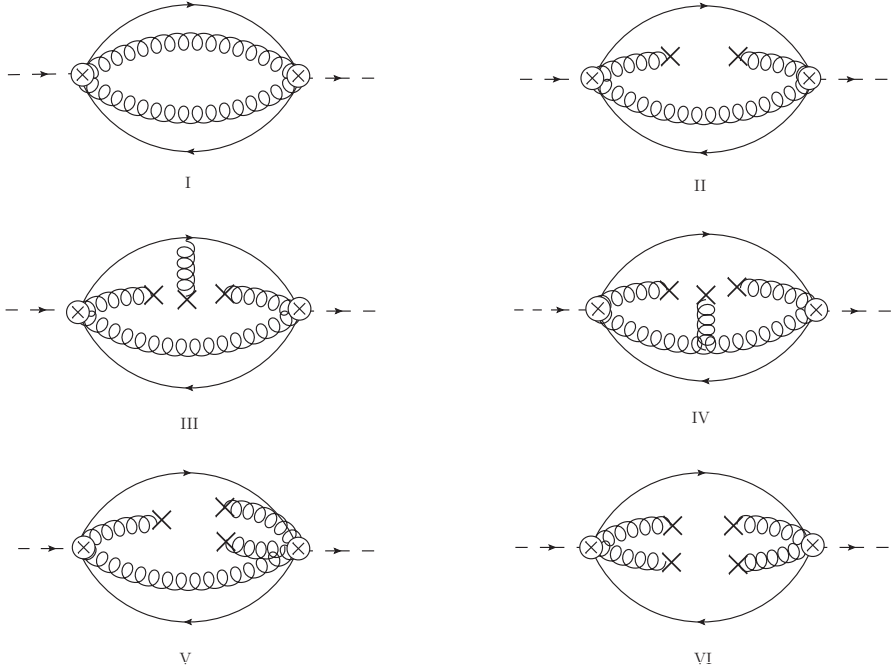}\\
  \caption{The typical LO Feynman diagrams of a double-gluon heavy quarkonium hybrid state that contribute to the spectral densities in Eq.\eqref{Pi-OPE-Expand}, where the permutation diagrams are implied. Digram I represents the contribution from perturbative part, and diagrams II, III-V, and VI denote the two-gluon condensate, tri-gluon condensates, and four-gluon condensate, respectively. We note that diagram I is proportional to $\alpha_s^2\times g_s^0$, diagrams II-VI are proportional to $\alpha_s^2 \times g_s^2$, respectively. So all the diagrams up to four-gluon condensate at leading order of $\alpha_s$ are depicted.}\label{Feyn-Diag}
\end{figure}

On the QCD side of QCD sum rules, based on the dispersion relation, the correlation function $\Pi(q^2)$ can be expressed as follows:
\begin{eqnarray}\label{Pi-OPE}
  \Pi^{\text{QCD}} (q^2) = \int_{4 m_Q^2}^\infty ds \frac{\rho^{\text{OPE}}(s)}{s - q^2},
\end{eqnarray}
where $\rho^{\text{OPE}}(s) = \text{Im} [\Pi^{\text{OPE}}(s)]/\pi$, and
\begin{eqnarray}\label{Pi-OPE-Expand}
  \rho^{\text{OPE}}(s) &=& \rho^{\text{pert}}(s) + \rho^{\langle G^2 \rangle}(s) + \rho^{\langle G^3 \rangle}(s) + \rho^{\langle G^4 \rangle}(s),
\end{eqnarray}
where $\rho^{\text{pert}}(s)$, $\rho^{\langle G^2 \rangle}(s)$, $\rho^{\langle G^3 \rangle}(s)$, and $\rho^{\langle G^4 \rangle}(s)$ denote the spectral densities of the perturbative part, the two-gluon condensate contribution, the tri-gluon condensate contribution, and the four-gluon condensate contribution, respectively. For instance, to calculate the perturbative part $\rho^{\text{pert}}(s)$, we firstly combine two full propagators of heavy quarks given in Eq.\eqref{prop-heavy} and two full propagators of the gluons shown in Eqs.(\ref{pert-gluon},\ref{gluon-propagator-single-gluon}), and then choose the perturbative term which does not contain any condensate terms. Eventually, we utilize the technique explicitly shown in Refs.\cite{Reinders:1984sr,Albuquerque:2012jbz} to calculate $\rho^{\text{pert}}(s)$. The same procedure is applicable to the calculations of other spectral densities that contain gluon condensates. The typical LO Feynman diagrams of a double-gluon heavy quarkonium hybrid state that contribute to the spectral densities in Eq.\eqref{Pi-OPE-Expand} are shown in Fig.~\ref{Feyn-Diag}, where diagram I represents the contribution from perturbative part, and diagrams II, III-V, and VI denote the two-gluon condensate, tri-gluon condensates, and four-gluon condensate, respectively. We note from Fig.\ref{Feyn-Diag} that diagram I is proportional to $\alpha_s^2\times g_s^0$, diagrams II-VI are proportional to $\alpha_s^2 \times g_s^2$, respectively. Note that the permutation diagrams are implied in Fig.\ref{Feyn-Diag}, so all the diagrams up to four-gluon condensate at leading order of $\alpha_s$ are depicted and calculated in our work. The lengthy expressions of spectral densities in Eq.\eqref{Pi-OPE-Expand} are deferred to the Appendix.

On the phenomenological side of QCD sum rules, the spectral funtion $\rho^{\text{phen}}(s)$ is defined using the pole plus continuum approximation
\begin{eqnarray}
  \rho^{\text{phen}}(s) = \lambda_{H_Q} ^2 \delta(s - M_{H_Q}^2) + \rho^{h}(s),
\end{eqnarray}
where the subscript $H_Q$ ($Q = c$ or $b$) denotes the lowest lying hybrid state, $M_{H_Q}$ represents its mass, $\rho^h(s)$ means the spectral density which includes the contributions from higher excited states and the continuum states above the threshold $s_0$. The coupling constant $\lambda_H$ is defined by $\langle 0|j_\mu | H_{Q} \rangle = \lambda_{H_Q} \epsilon_\mu$.

After isolating the ground state contribution from the hybrid state, we obtain the correlation function $\Pi^{\text{phen}}(q^2)$ in dispersion integral over the physical region, \emph{i.e.},
\begin{eqnarray}\label{Pi-phen}
  \Pi^{\text{phen}}(q^2)= \frac{\lambda_{H_Q}^2}{(M_{H_Q})^2 - q^2} + \int_{s_0}^{\infty} d s \frac{\rho^h(s)}{s-q^2}.
\end{eqnarray}

For extracting reliable results from the comparison between the two representations of the correlation function, one should guarantee a good OPE convergence on the QCD side and simultaneously suppress the contributions from higher excited states and the continuum states on the phenomenological side. A practical way of doing this is to utilize the Borel transformation, whose definition
is given by:
\begin{equation}	
  {\cal B}\! \left[ \Pi(Q^2) \right] \equiv \Pi(M_B^2) =
  \lim\limits_{\tiny \begin{matrix} Q^2, n \rightarrow \infty \\ Q^2/n = M_B^2 \end{matrix}}
  \frac{(-1)^n (Q^2)^{n+1}}{n!} \left( \frac{\partial}{\partial Q^2} \right)^n \!\Pi(Q^2)
\end{equation}
where $Q^2$ is the four-momentum of the particle in the Euclidean space $(Q^2 = -q^2)$,
and $M_B^2$ is a free parameter of the sum rule.

Performing Borel transformation on the QCD side Eq.~\eqref{Pi-OPE} and the phenomenological side Eq.~\eqref{Pi-phen}, and using quark-hadron duality,
we can establish the main function of QCD sum rules, that is:
\begin{eqnarray}\label{main-function}
  \int_{4m_c^2}^{s_0} \rho^{\text{OPE}}(s) e^{-s/M_B^2} ds = \lambda_{H_Q}^2 e^{-M_{H_Q}^2/M_B^2},
\end{eqnarray}
where the so-called quark-hadron duality approximation~\cite{P.Col} is used which has the following form:
\begin{eqnarray}
  \int_{s_0}^\infty \rho^{\text{OPE}}(s) e^{-s/M_B^2} ds \simeq \int_{s_0}^\infty \rho^{h}(s) e^{-s/M_B^2} ds.
\end{eqnarray}

Then we can extract the mass of the hybrid state from the main function \eqref{main-function}, which reads:
\begin{eqnarray}
  M_{H_Q}^{i}(s_0, M_B^2) &=& \sqrt{-\frac{L_1(s_0, M_B^2)}{L_0(s_0, M_B^2)}}, \label{mass-function}
\end{eqnarray}
where the superscript $i$ runs from $A$ to $D$, respectively. The moments $L_1$ and $L_0$ are, respectively, defined as
\begin{eqnarray}
  L_0(s_0, M_B^2) &=& \int_{4 m_Q^2}^\infty ds \, \rho^{\text{OPE}}(s) e^{-s/M_B^2}, \label{L0} \\
  L_1(s_0, M_B^2) &=& \frac{\partial}{\partial (M_B^2)^{-1}} L_0(s_0, M_B^2).
\end{eqnarray}

\section{Numerical analyses}\label{Numerical}
For numerical evaluation, the leading order strong coupling constant
\begin{eqnarray}
\alpha_{s}(M_{B}^{2}) = \frac{4\pi}{(11-\frac{2}{3}n_{f}) \ln(\frac{M_{B}^{2}}{\Lambda_{\text{QCD}}^{2}})},
\end{eqnarray}
is adopted with $\Lambda_{\text{QCD}} = 300$ MeV and $n_f$ being the number of active quarks~\cite{Wan:2020fsk}. Additionally, in order to yield meaningful physical results in QCD sum rules, as in any practical theory, one needs to give certain inputs. These input parameters are taken from~\cite{ParticleDataGroup:2020ssz, Narison:2011xe, Narison:2018dcr}, whose explicit values read:
\begin{eqnarray}\label{inputparamters}
m_c(m_c)&=&\bar{m}_c=(1.27 \pm 0.02) \text{GeV}, m_b(m_b) = \bar{m}_b=(4.18^{+0.03}_{-0.02}) \text{GeV}, \nonumber \\
\langle \alpha_s G^2 \rangle &=& (6.35 \pm 0.35)\times 10^{-2} \; \text{GeV}^{4},\;\;\langle g_s^3 G^3 \rangle = (8.2 \pm 1.0) \times \langle \alpha_s G^2\rangle \text{GeV}^2,
\end{eqnarray}
where we use the ``running masses" for the heavy quarks in the $\overline{\text{MS}}$ scheme. It is important to note that the vacuum saturation approximation is used in this work in the calculation of $\langle G^4 \rangle$ contribution~\cite{Shifman:1980ui, Bagan:1984zt}. In order to take into account the error due to the violation of the vacuum approximation, we can introduce a parameter $\kappa$,
\begin{eqnarray}
  \langle \alpha_s G^2 \rangle^2 \to \kappa \langle \alpha_s G^2 \rangle^2,
\end{eqnarray}
the value $\kappa =1$ stands for the vacuum saturation approximation, while the value $\kappa \neq 1$ parameterizes its violation. We consider the result obtained by using the factorized $\langle G^4 \rangle$ as the central value ($\kappa =1$), and consider the variation due to the violation of the vacuum dominance (by a factor of $\kappa =2$) as a source of errors.

In establishing the QCD sum rules, there are two additional parameters $s_0$ and $M_B^2$ represented the threshold parameter and the Borel parameter, respectively. For a given $s_0$, the Borel parameter $M_B^2$ will be constrained by two criteria~\cite{P.Col, Tang:2019nwv}.
First, in order to extract the information on the ground state of the double-gluon heavy hybrid state, one should guarantee pole contribution (PC) is larger than 40\%, which can be formulated as
\begin{eqnarray}\label{RatioPC}
  R_{i}^{\text{PC}} = \frac{L_0(s_0, M_B^2)}{L_0(\infty, M_B^2)} \; ,
\end{eqnarray}
where the subscript $i$ runs from A to D. Under this constraint, the contribution of higher excited and continuum states will be suppressed. This criterion gives rise to a critical value of $M_B^2$, which is the upper limit of $M_B^2$ nominated as $(M_B^2)_{max}$.

To insure the convergence of Eq.\eqref{L0}, we should require an OPE series decreasing order by order, for $\kappa =1$ and 2, respectively. Then, one can determine another critical value of $M_B^2$ from the ratios of various terms in Eq.\eqref{L0} to the entire moment $L_0(s_0, M_B^2)$, defined as
\begin{eqnarray}\label{Ratiodim}
  R_{i}^{\text{cond}} = \frac{L_0^{\text{cond}}(s_0, M_B^2)}{L_0(s_0, M_B^2)}\, ,
\end{eqnarray}
which corresponds to the lower limit of $M_B^2$ called $(M_B^2)_{min}$. Here, the subscript $i$ runs from $A$ to $D$, and the superscript `cond' denotes the perturbative term and different condensate terms in Eq.\eqref{L0}, respectively. As a consequence, we obtain the proper Borel window of $M_B^2$ for a given $s_0$, which is the region between $(M_B^2)_{min}$ and $(M_B^2)_{max}$.

In practice, to know whether the OPE convergence is satisfied, we firstly restrict that the highest condensate contribution, $\langle G^4 \rangle$, should be less than 15\% and 25\% of the total OPE side for $\kappa =1$ and $\kappa=2$, respectively. Then, we can select the one which has an OPE series decreasing order by order.

It is obvious that the Borel window depends on the threshold parameter $s_0$. Therefore, we need to vary the value of $s_0$ in a possible region, until we find an optimal value of $s_0$ which corresponds to a smooth plateau for the hybrid mass $M_{H_Q}$ in its Borel window given by the two criteria mentioned above. On the smooth plateau, the hybrid mass $M_{H_Q}$ should be in principle independent of the Borel parameter $M_B^2$, or at least only shows weak dependence.

\begin{figure}[htb]
\begin{center}
\includegraphics[width=7.6cm]{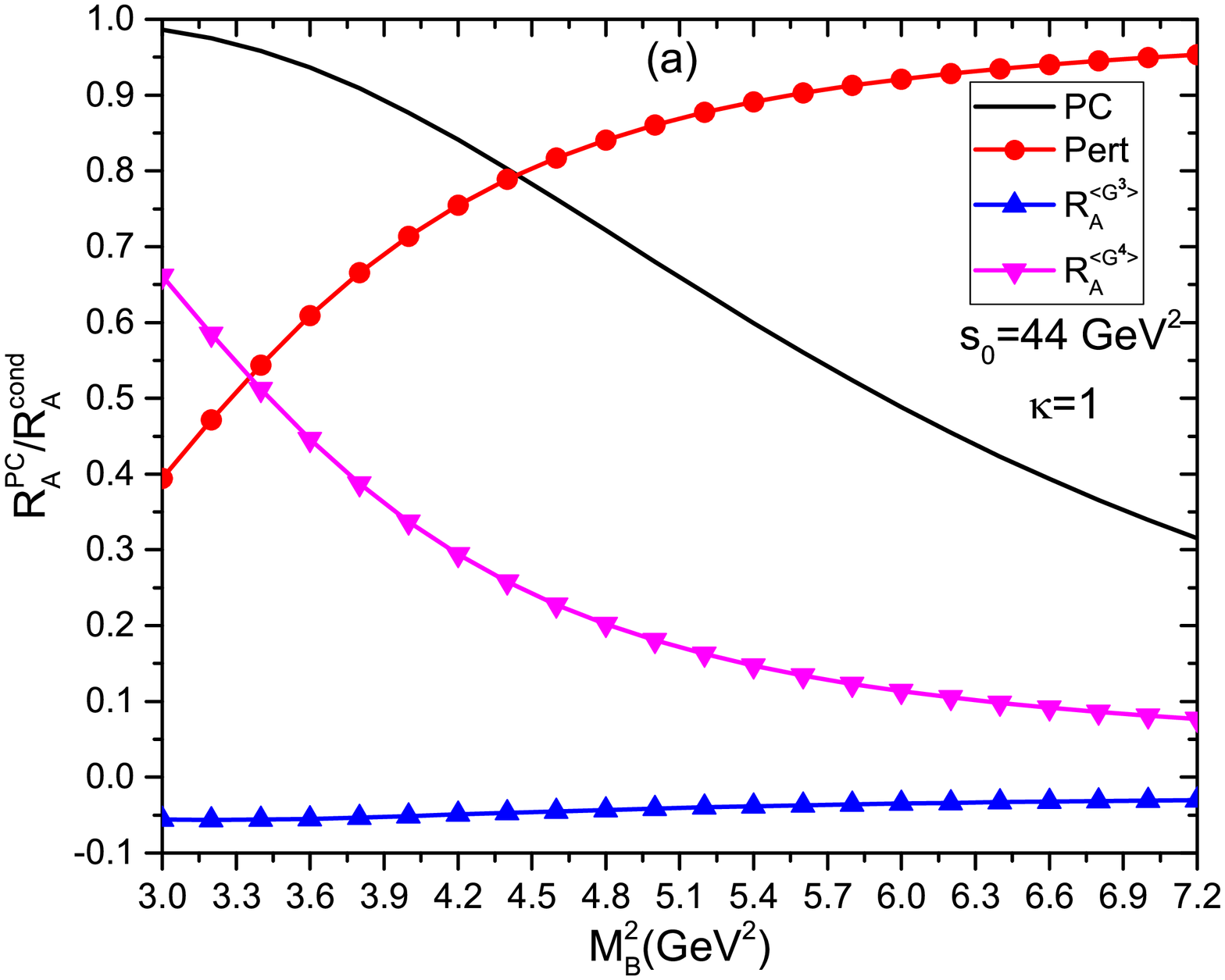}
\includegraphics[width=7.6cm]{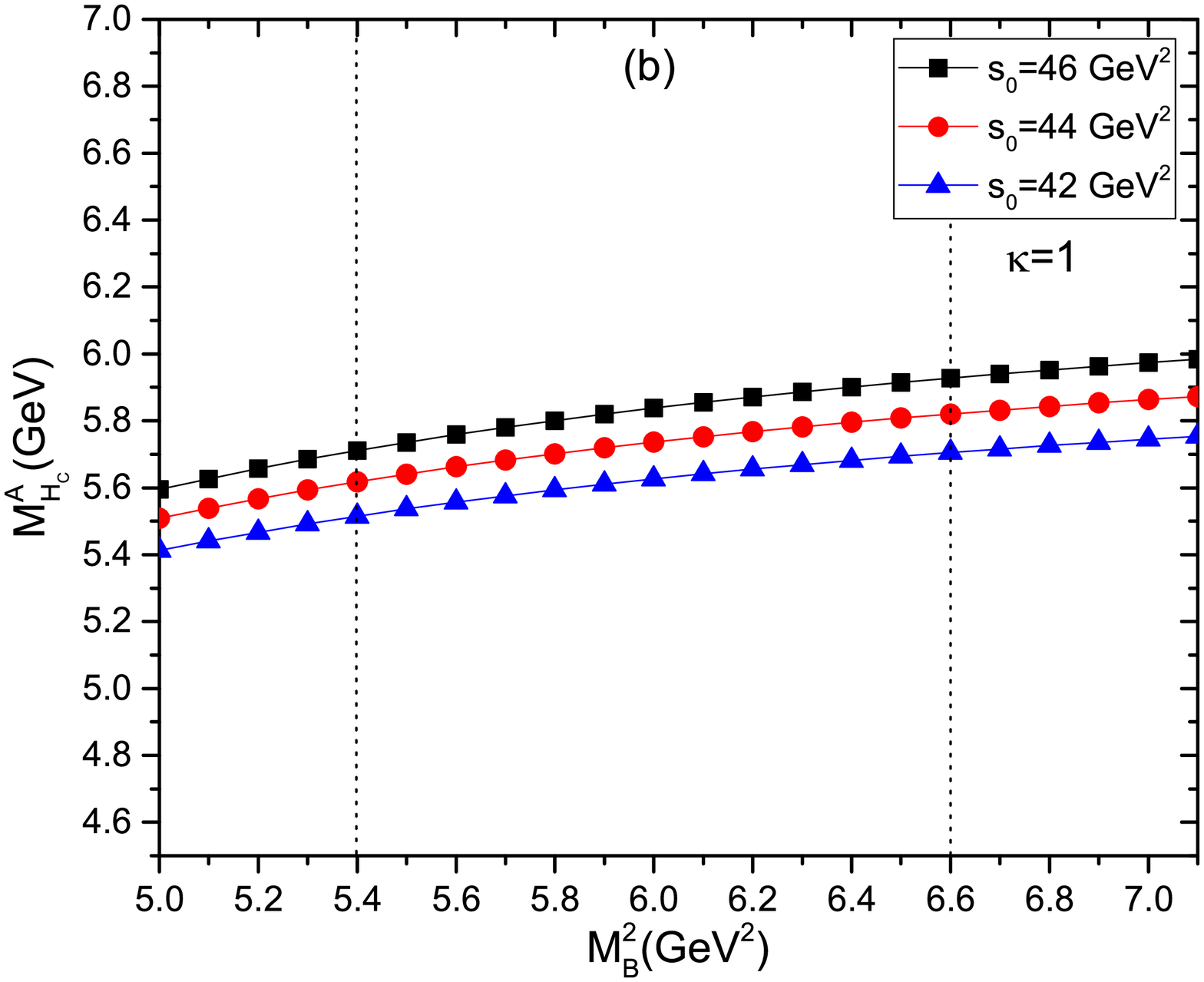}
\caption{(color). The figures for current A with $\kappa =1$. (a) The pole contribution ratio $R_A^{\text{PC}}$ and OPE convergence ratio $R_A^{\text{cond}}$ as functions of the Borel parameter $M_B^2$ with the central value of $s_{0}$; (b) The mass $M_{H_c}^A$ as a function of $M_B^2$ for $s_0= 42\, \text{GeV}^2$, $44\, \text{GeV}^2$, and $46 \, \text{GeV}^2$ from down to up, respectively, and the two vertical lines indicate the upper and lower bounds of the proper Borel window with the central value of $s_{0}$.}
\label{fig2}
\end{center}
\end{figure}

\begin{figure}[htb]
\begin{center}
\includegraphics[width=7.6cm]{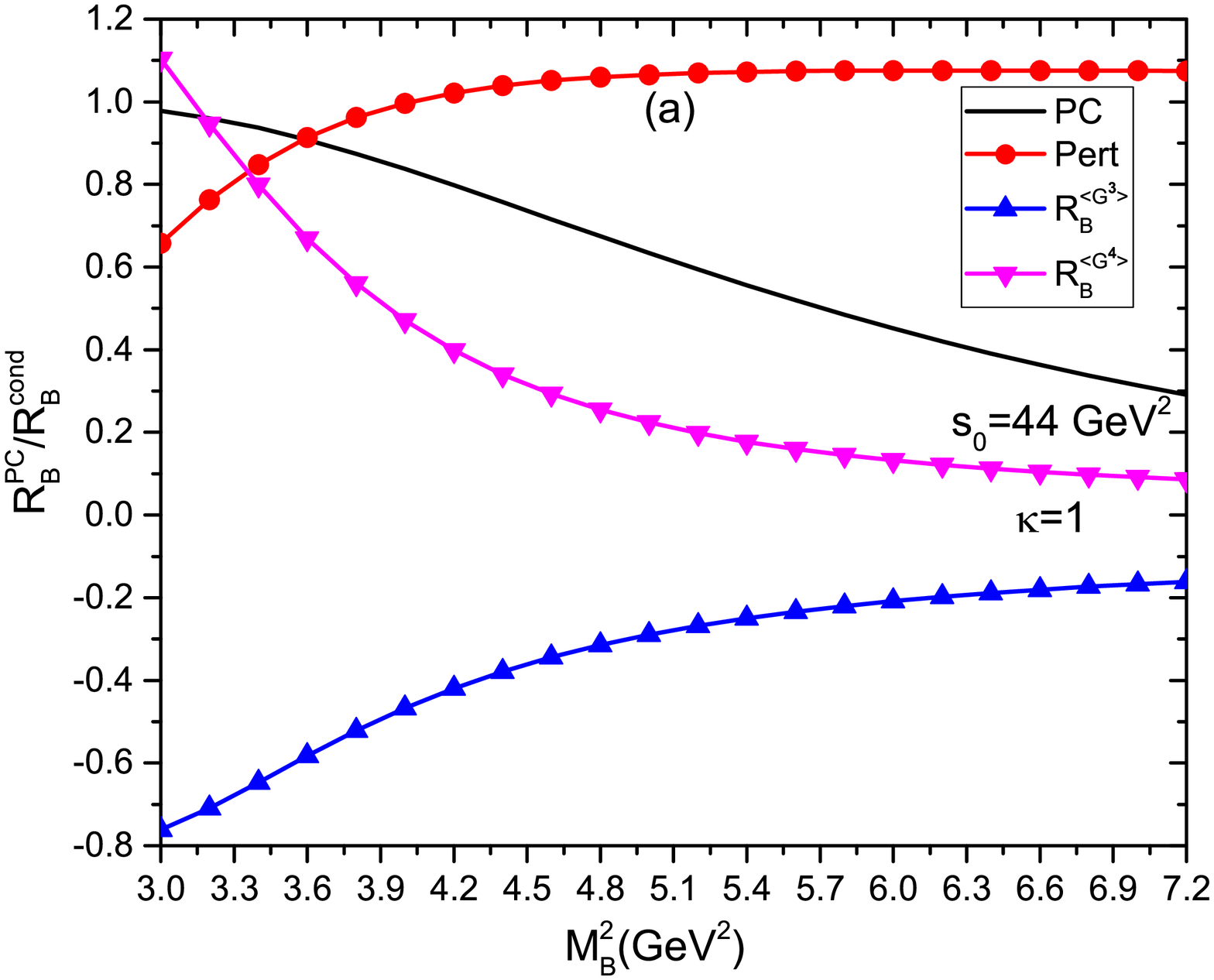}
\includegraphics[width=7.6cm]{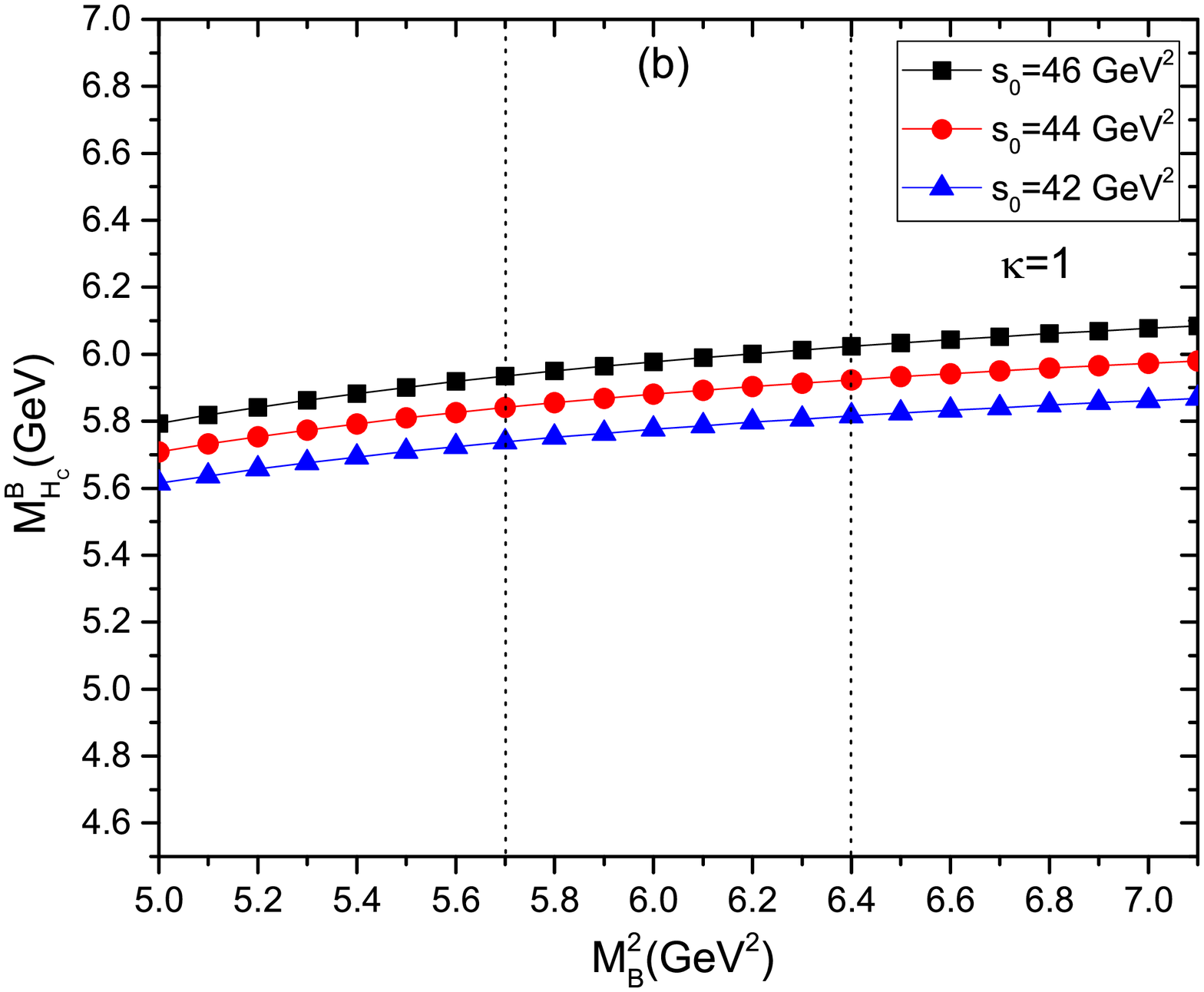}
\caption{(color). The same caption as in Fig.~\ref{fig2}, but for current B.}
\label{fig3}
\end{center}
\end{figure}

\begin{figure}[htb]
\begin{center}
\includegraphics[width=7.6cm]{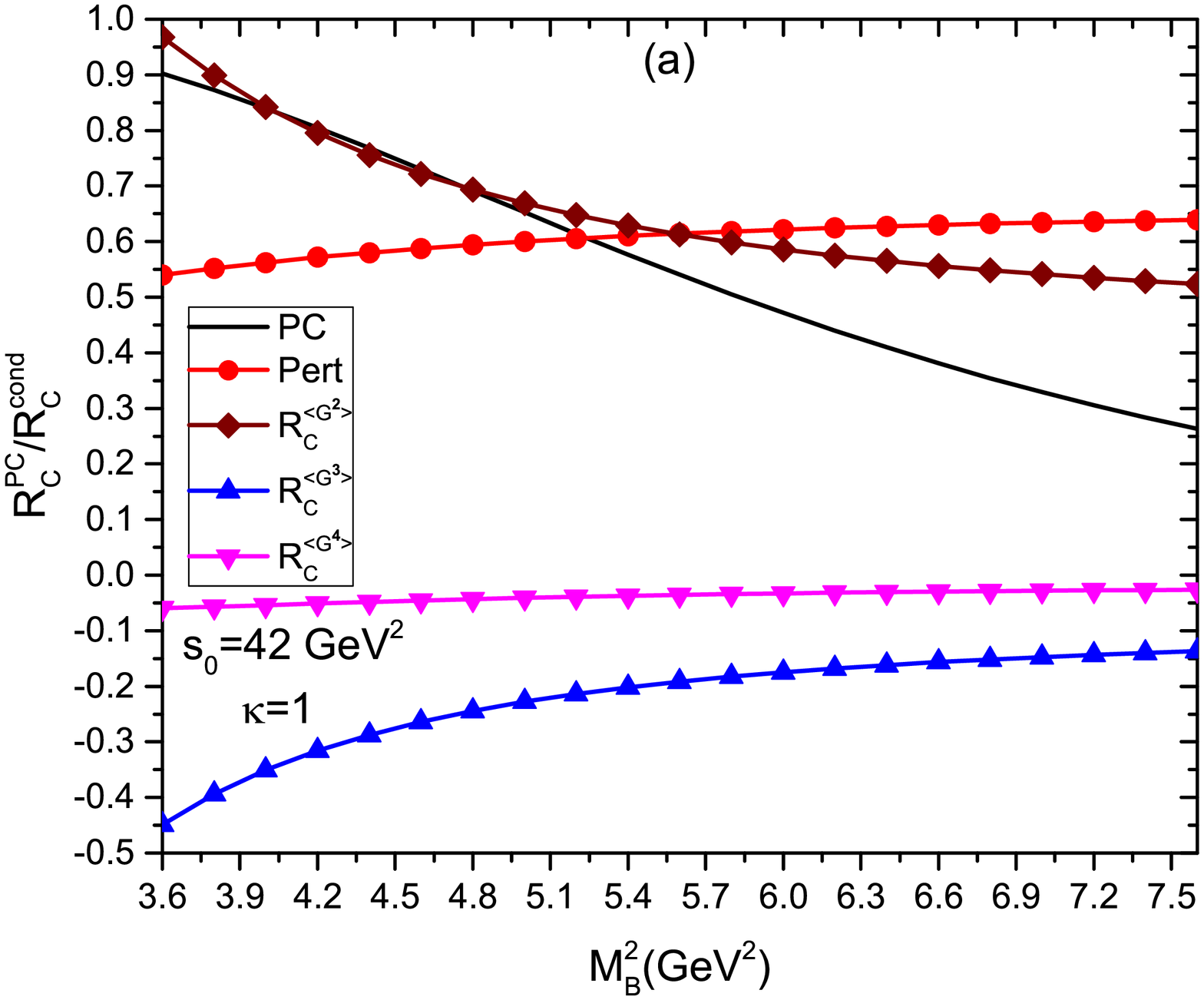}
\includegraphics[width=7.6cm]{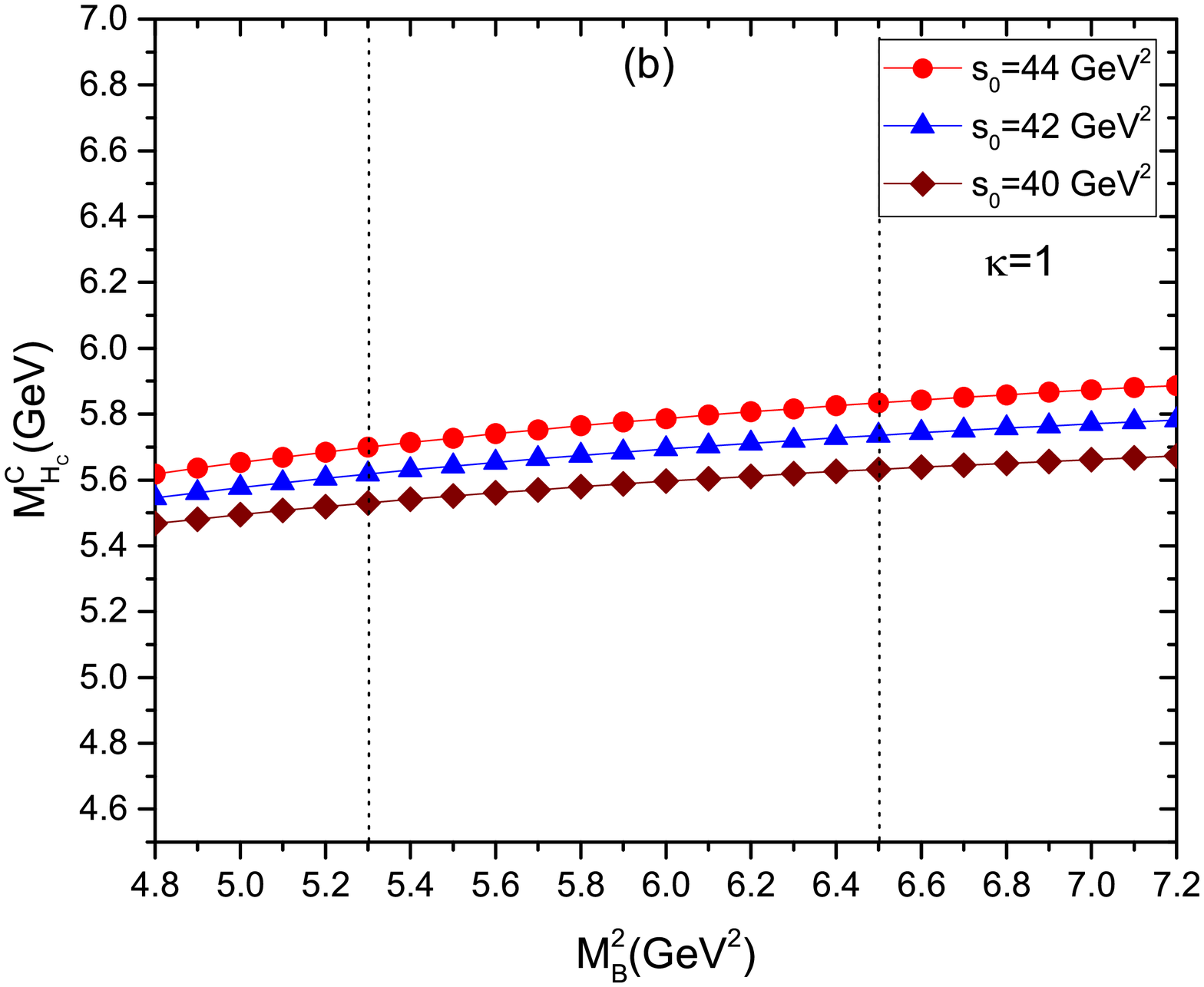}
\caption{(color). The same caption as in Fig.~\ref{fig2}, but for current C.}
\label{fig4}
\end{center}
\end{figure}

For case A with $\kappa =1$, we plot the two ratios $R_A^{\text{PC}}$ and $R_A^{\text{cond}}$ as functions of the Borel parameter $M_B^2$ in Fig.~\ref{fig2}(a) at the proper value $s_0 = 44 \, \text{GeV}^2$, and the mass curves as functions of $M_B^2$ in Fig.~\ref{fig2}(b). Two vertical lines in Fig.~\ref{fig2}(b) indicate the upper and lower bounds of the proper Borel window for the central value of $s_0$, and the so-called stable plateau between these two vertical lines exists, where the proper Borel window refers to the one that fulfills the constraint $R_A^{\langle G^4 \rangle}<15\%$. To estimate the uncertainty introduced by $s_0$, we tentatively assign a $2\, \text{GeV}^2$ fluctuation from the optimal value $s_0 = 44\, \text{GeV}^2$, as shown in Fig.~\ref{fig2}(b). A similar situation happens for case B with $\kappa =1$, shown in Fig.~(\ref{fig3}). For case C, since the tentative restriction $R_C^{\langle G^4 \rangle}<15\%$ is satisfied in a wide range of the Borel parameter, as shown in Fig.~\ref{fig4}(a), the lower limit of $M_B^2$ is fixed by the requirement that the ratio $R_C^{\text{Pert}}$ is lager than 60\%. The mass figures in Figs.~(\ref{fig3}(b),\ref{fig4}(b)) also exhibit stable plateau within their proper Borel windows, respectively. However, for case D with $\kappa =1$, we find that no matter what value of $s_0$ and $M_B^2$ are taken to be, no proper Borel window for a stable plateau exists. That means the current structure in Eq.~\eqref{currentD} does not support the corresponding hybrid.

The resulting windows of the Borel parameter $M_B^2$, threshold values $s_0$, pole contributions (PC), two-gluon contributions, tri-gluon contributions, four-gluon contributions for cases A, B, and C with $\kappa =1$ are shown explicitly in Table.~\ref{tab1}, respectively. From Table.~\ref{tab1}, we can see that, for case A, although the pole dominance of the phenomenological side is well satisfied in the proper Borel window, the OPE convergence constraint is violated due to $|R_A^{\langle G^4\rangle}/R_A^{\langle G^3\rangle}| >1$. Hence, we exclude case A when we make further numerical analyses in the following text.

\begin{table}[htb]
\begin{center}
\begin{tabular}{|c|c|c|c|c|c|c|c|c}\hline\hline
 $\kappa =1$ & $M_B^2 (\rm{GeV}^2)$  & $s_{0} (\rm{GeV^{2}})$ & PC & Pert &  $R_{i}^{\langle G^2 \rangle}$   &  $R_{i}^{\langle G^3 \rangle}$ &  $R_{i}^{\langle G^4 \rangle}$  \\ \hline
 case A & $5.40\!-\!6.60$ & 44 & $(60\!-\!40)\%$  & $(89\!-\!94)\%$ & 0   & $[(-3.83)\!-\!(-3.20)]\%$  & $(14.72\!-\!9.17)\%$   \\ \hline
 case B & $5.70\!-\!6.40$ & 44  & $(50\!-\!40)\%$ & $(107\!-\!108)\%$ & 0 & $[(-23)\!-\!(-19)]\%$  & $(15.20\!-\!11.27)\%$  \\ \hline
case C & $5.30\!-\!6.50$ & 42 & $(60\!-\!40)\%$ & $(61\!-\!63)\%$ & $(64\!-\!56)\%$  & $[(-21)\!-\!(-16)]\%$  & $[(-3.80)\!-\!(-3.02)]\%$  \\ \hline
 \hline
\end{tabular}
\end{center}
\caption{$\kappa=1$. The windows of the Borel parameter $M_B^2$, threshold parameters $s_0$, pole contributions, two-gluon contributions, tri-gluon contributions, and four-gluon contributions of $\bar{c} g g c$ hybrid states for cases A, B, and C, respectively.}
\label{tab1}
\end{table}

As already mentioned, we need to consider the variation due to the violation of the vacuum dominance (by a factor of $\kappa=2$) as a source of errors. Therefore, we list the resulting windows of the Borel parameter $M_B^2$, threshold values $s_0$, pole contributions (PC), two-gluon contributions, tri-gluon contributions, four-gluon contributions for cases B and C with $\kappa =2$ in Table.\ref{tab2}, respectively.

\begin{table}[htb]
\begin{center}
\begin{tabular}{|c|c|c|c|c|c|c|c|c}\hline\hline
 $\kappa=2$ & $M_B^2 (\rm{GeV}^2)$  & $s_{0} (\rm{GeV^{2}})$ & PC & Pert &  $R_{i}^{\langle G^2 \rangle}$   &  $R_{i}^{\langle G^3 \rangle}$ &  $R_{i}^{\langle G^4 \rangle}$  \\ \hline
 case B & $5.80\!-\!6.50$ & 44  & $(52\!-\!40)\%$ & $(94\!-\!97)\%$ & 0 & $[(-19)\!-\!(-17)]\%$  & $(25\!-\!20)\%$  \\ \hline
case C & $5.40\!-\!6.40$ & 42 & $(57\!-\!40)\%$ & $(63\!-\!65)\%$ & $(65\!-\!58)\%$  & $[(-21)\!-\!(-17)]\%$  & $[(-7.72)\!-\!(-6.32)]\%$  \\ \hline
 \hline
\end{tabular}
\end{center}
\caption{The same caption as in Table.\ref{tab1}, but for cases B and C with $\kappa=2$.}
\label{tab2}
\end{table}

From Table.~\ref{tab2}, we find that, for case B, the OPE convergence constraint is violated for $\kappa =2$ because of $|R_A^{\langle G^4\rangle}/R_A^{\langle G^3\rangle}| >1$. Hence, we also exclude case B in the following numerical analyses. Ultimately, we conclude that both the pole dominance of the phenomenological side and the OPE convergence are well satisfied for case C.

\begin{figure}[htb]
\begin{center}
\includegraphics[width=12cm]{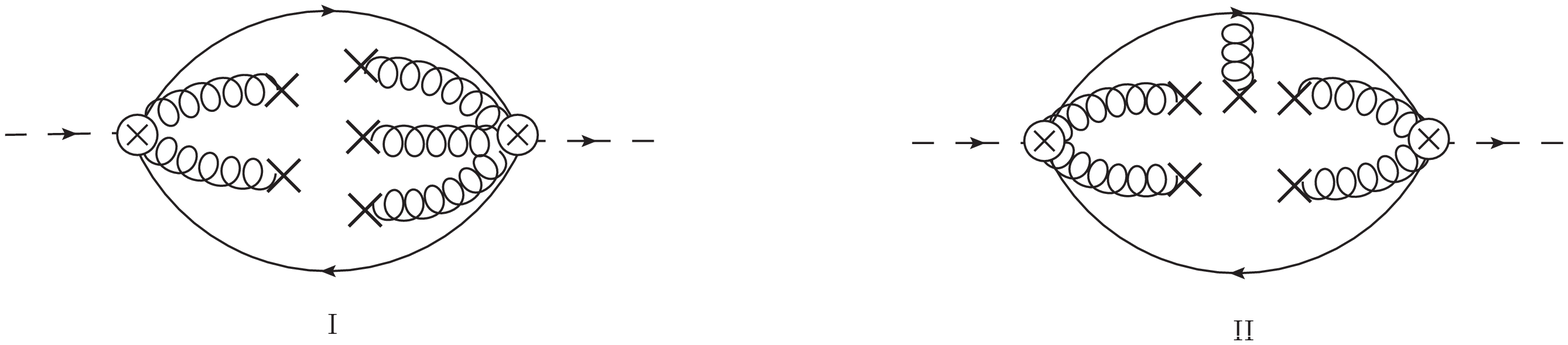}
\caption{The typical LO Feynman diagrams that contribute to $\langle G^5 \rangle$ term, where the permutation diagrams are implied.}
\label{figG5}
\end{center}
\end{figure}

For case C, to safely neglect the contribution from $d \geq 10$, where $d$ represents the dimension of the condensate term, it is necessary to guarantee $|R_C^{\langle G^5 \rangle}/R_C^{\langle G^4 \rangle}<1|$. To this end, we should calculate the contribution for $\langle G^5 \rangle$ condensate ($d=10$), then test whether the size of the $d=10$ term is smaller than that from the $d=8$ term in the present Borel window listed in Table.~\ref{tab1}. The leading order Feynman diagrams of the $d=10$ term are depicted in Fig.~\ref{figG5}, where the permutation diagrams are implied. We put the details on the calculation and analytic expression of the $d=10$ term in the Appendix.

\begin{figure}[htb]
\begin{center}
\includegraphics[width=8cm]{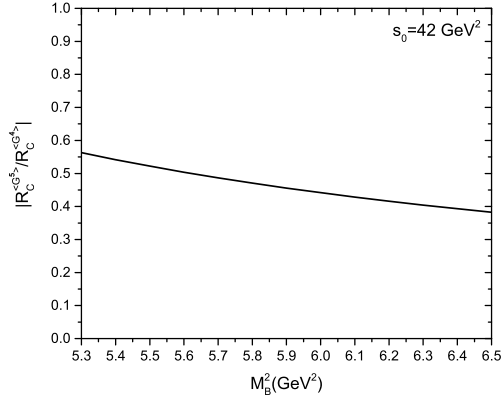}
\caption{$\kappa =1$. The ratio $|R_C^{\langle G^5\rangle}/R_C^{\langle G^4 \rangle}|$ as a function of the Borel parameter $M_B^2$ in the valid Borel window with the central value of $s_0$.}
\label{figRatioG5C}
\end{center}
\end{figure}

From Fig.~\ref{figRatioG5C}, we can conclude that the condensate contribution from the $d=10$ term is much less than the $d=8$ term in the present Borel window for $\kappa =1$, and we can safely neglect it. Then, it is valid to truncate the OPE at $d=8$, and the present Borel window is the valid Borel window that satisfies all the constraints in the QCD sum rules. Moreover, since the value of $|R_C^{\langle G^5\rangle}/R_C^{\langle G^4 \rangle}|$ will decrease as the parameter $\kappa$ increases, we can obtain a better OPE convergence for $\kappa =2$. Therefore, we can make a reliable mass prediction for case C.

Now, we can determine the masses of the vector double-gluon charmonium hybrid state for current C with $\kappa =1$ and $\kappa =2$, which are summarized in Table.\ref{tab3}, where the subscript $H_c$ denotes the hybrid state in the c-quark sector; the error bars stem from the uncertainties of the Borel parameter $M_B^2$, the threshold parameter $s_0$, the condensate parameters $\langle g_s^2 G^2\rangle$ and $\langle g_s^3 G^3\rangle$, and the quark mass $\bar{m}_c$ listed in Eq.\eqref{inputparamters}. It should be noted that the variations of the Borel window in the region of $(s_0^{\text{center}}\pm 2) \, \text{GeV}^2$ have been considered in our estimation of the uncertainties, where $s_0^{\text{center}}$ represents the central value of $s_0$.

\begin{table}[h]
\begin{tabular}{|c|c|c|}\hline\hline
 & $\kappa$=1 & $\kappa$=2 \\ \hline
case C & $M^{C,1^{--}}_{H_c} = (5.68^{+0.20}_{-0.35})\, \text{GeV} \; ,$ & $M^{C,1^{--}}_{H_c} = (5.71^{+0.19}_{-0.15})\, \text{GeV} \; ,$  \\ \hline
 \hline
\end{tabular}
\caption{Mass predictions for the $1^{--}$ double-gluon charmonium hybrids both with $\kappa=1$ and $\kappa =2$, respectively. The error bars are obtained by taking into account the uncertainties of the Borel windows, $s_0$, and the input parameters shown in Eq.\eqref{inputparamters}.}
\label{tab3}
\end{table}

Eventually, by considering all the uncertainties mentioned above, we obtain the mass prediction of the $1^{--}$ double-gluon charmonium hybrid state, which is
\begin{eqnarray}
M^{C,1^{--}}_{H_c} &=& (5.68^{+0.22}_{-0.35}) \, \text{GeV} \; , \label{eq-mass-3}
\end{eqnarray}
and find that it is in the region of $5.33\, \text{GeV} < M_{H_c}<5.90\, \text{GeV}$.

By replacing the mass of the charm quark with the the bottom quark in Eq.\eqref{Pi-OPE} and performing the same numerical analyses, we can obtain the corresponding prediction for the $1^{--}$ double-gluon bottomonium hybrid state, whose mass is
\begin{eqnarray}
M^{C,1^{--}}_{H_b} &=& (11.51^{+0.17}_{-0.31}) \, \text{GeV} \; , \label{eq-mass-6}
\end{eqnarray}
respectively, where the subscript $H_b$ represents the hybrid state in b-quark sector. By including the uncertainties of this bottomonium hybrid mass, we find that it is in the range of $11.20 \sim 11.68\, \text{GeV}$.

\section{Decay analyses}

\begin{figure}[htb]
\begin{center}
\includegraphics[width=13cm]{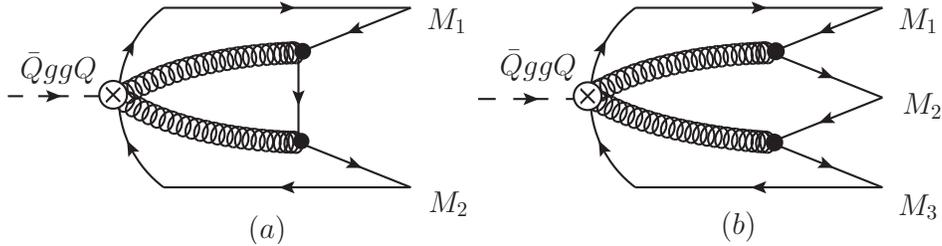}
\caption{Two possible decay processes of the double-gluon heavy quarkonium hybrids $\bar{Q}ggQ$, where the final states are represented as $M_1$, $M_2$, and $M_3$. The same figures have been given in Ref.~\cite{Chen:2021smz} for explaining the possible decay processes of the double-gluon hybrids in light quark sector.}
\label{fig5}
\end{center}
\end{figure}

\begin{table}[hbt]
\caption{Some possible two- and three-meson decay channels of the $\bar{c}ggc$ hybrids with the quantum number $I^G(J^{PC}) = 0^-(1^{--})$ that is consistent with the $Y$ states, where we only keep the channels up to P-wave decays. Here, for brevity, the notation $D^{(*)} \bar{D}^{(*)}$ denotes both $D \bar{D}$ and $D^{*} \bar{D}^{*}$, and the notation $D^* \bar{D}$ represents not only $D^* \bar{D}$ but also $D \bar{D}^*$.}
\begin{tabular}{|c|c|c|}\hline
\hline
                   & S-wave   & P-wave   \\ \hline
\multirow{1}{*}{Two-meson}& ---  & $D^{(*)} \bar{D}^{(*)}$, $D_s^{(*)} \bar{D}_s^{(*)}, D^* \bar{D}, D_s^* \bar{D}_s$  \\ \hline
\multirow{3}{*}{Three-meson} & $D^* \bar{D}^* \pi, D^* \bar{D}^* \eta, D^{(*)} \bar{D}^{(*)} \rho,$   & $D^{(*)} \bar{D}^{(*)} h_1, D^{(*)}\bar{D}^{(*)} b_1, D^{(*)} \bar{D}^{(*)} a_{0,1,2},$  \\
& $ D^{(*)} \bar{D}^{(*)} \omega, D^* \bar{D}\pi, D^* \bar{D}\rho,$  & $D^{(*)} \bar{D}^{(*)} f_{0,1,2}, D^* \bar{D}h_1, D^* \bar{D}b_1, $  \\ & $D^* \bar{D} \omega, D^* \bar{D}\eta$ & $D^* \bar{D} a_{0,1,2}, D^* \bar{D} f_{0,1,2}$
\\ \hline
\hline
\end{tabular}\label{tableII}
\end{table}

As shown in Fig.~\ref{fig5}, the double-gluon heavy quarkonium hybrids can decay into a pair of charmed/bottomed mesons or a pair of charmed/bottomed mesons together with a light meson by exciting two light quark ($u$, $d$, or $s$) pairs from the two valence gluons. It should be noted that these two possible decay modes are both at $\cal{O}$$(\alpha_s)$ order, though they are OZI-allowed processes.

As shown in Table.~\ref{tableII}, apart from the S-wave decays in the two-meson decay patterns which violate the conservation of the parity, there exist P-wave decays in the two-meson decay patterns, and both S-wave and P-wave decays in the three-meson decay patterns. In order to select some better decay channels, for a qualitative analysis, we only consider two aspects that affect the decay branching ratios of these predicted hybrids: the phase space factor and the P-wave suppression. In view of these two aspects, we notice that, the P-wave two-meson decay pattern has a bigger phase factor than the three-meson case, whereas, it is suppressed by the excited energy corresponding to the P-wave interaction between its final states; the S-wave three-meson decay pattern does not need the excited energy of the P-wave interaction, but has a smaller phase factor compared to the two-meson decay pattern. Therefore, each type of these decay channels has an advantage and a disadvantage. These behaviors will be useful for identifying the nature of the double-gluon heavy quarkonium hybrids.

Amongst them listed in Table.~\ref{tableII}, we suggest the decay channels $H_c \to D \bar{D}/D^* \bar{D}/D^{*} \bar{D}^{*}$ with P wave and $H_c \to D^* \bar{D}^{*} \pi/D^* \bar{D}^*\eta/D \bar{D}\rho/D \bar{D}\omega$ with S wave as the accessible decay channels for the double-gluon charmonium hybrids, which are expected to be measured in Belle II, PANDA, Super-B, GlueX, and LHCb in the near future.

\section{Conclusions}
Since a series of newly observed `exotic' states in the charmonium energy region possess the quantum number $J^{PC} = 1^{--}$, which are nominated as $Y$ states. Therefore, it is reasonable to believe that there will exist heavier $Y$ states, which may be composed of one charm quark and one anti-charm quark together with two gluons. In this work, we firstly construct four currents of the vector ($J^{PC} = 1^{--}$) double-gluon charmonium ($\bar{c}ggc$) hybrid. Then, we utilize the method of QCD sum rules to evaluate their masses. We find that the mass of $\bar{c}ggc$ hybrid lies in $M_{H_{c}}$ = $5.33 \sim 5.90$ GeV, while in the bottom sector the mass of $\bar{b}ggb$ hybrid may be situated in $M_{H_b} = 11.20 \sim 11.68$ GeV. The contributions up to dimension eight at leading order of $\alpha_s$ (LO) in the operator product expansion are taken into account in our calculation.

We depict two possible decay processes of the double-gluon heavy quarkonium hybrids in Fig.~\ref{fig5} and list their allowed two- and three-meson decay channels in Talbe.~\ref{tableII}, where we keep the channels up to P-wave decays. As a result, we suggest the decay channels $H_c \to D \bar{D}/D^* \bar{D}/D^{*} \bar{D}^{*}$ with P wave and $H_c \to D^* \bar{D}^{*} \pi/D^* \bar{D}^*\eta/D \bar{D}\rho/D \bar{D}\omega$ with S wave as the accessible decay channels of the double-gluon charmonium hybrids, which are expected to be measured in Belle II, PANDA, Super-B, GlueX, and LHCb in the near future.

%%%%%%%%%%%%%%%%%%%%%%%%%%%%%%%%%%%%%%%%%%%%%%%%%%%%%%%%%%%%%%%%%%%%%%
\vspace{.7cm} {\bf Acknowledgments} \vspace{.3cm}

This work was supported in part by the Science Foundation of Hebei Normal University under Contract No. L2016B08.

%%%%%%%%%%%%%%%%%%%%%%%%%%%%%%%%%%%%%%%%%%%%%%%%%%%%%%%%%%%%%%%%%%%%%%%

\begin{widetext}

%\newpage

\appendix \label{appendix}

\textbf{Appendix}

In this appendix, we list the spectral densities $\rho^{\text{OPE}}(s)$ in Eq.(\ref{Pi-OPE-Expand}) for all currents shown in Eqs. (\ref{currentA}-\ref{currentD}).

For case A, the expressions are summarized as follows:
\begin{eqnarray}
\rho^{\text{pert}}_{{\text{A, I}}}(s) &=& \frac{g_{s}^{4}}{2^{9}\times 9 \pi^{6}} \int_{\alpha_{min}}^{\alpha_{max}} d\alpha\int_{\beta_{min}}^{\beta_{max}} d\beta \left\{ -\frac{F_{\alpha\beta}^{3}m_{c}^{2}s(-\alpha-\beta+1)^{4}}{\alpha^{3}\beta^{3}} \right.\nonumber\\ &+& \frac{F_{\alpha\beta}^{5}(-\alpha-\beta+1)^{2}(-24(-\alpha - \beta+1)-18(\alpha+\beta))}{40\alpha^{4}\beta^{4}}\nonumber\\&-& \frac{F_{\alpha\beta}^{4}(-\alpha-\beta+1)^{2} (-12\alpha\beta s(-\alpha-\beta+1)+2m_{c}^{2}(\alpha + \beta)(-\alpha-\beta+1))}{8\alpha^{4}\beta^{4}}\nonumber\\
 &-& \left. \frac{F_{\alpha\beta}^{4}(-\alpha-\beta+1)^{2}(m_{c}^{2}(-\alpha -\beta+1)^{2})}{8\alpha^{4} \beta^{4}}\right\},
\end{eqnarray}
\begin{eqnarray}
\rho^{\langle G^{2} \rangle }_{{\text{A, II}}}(s) &=&0, \\
\rho^{\langle G^{3} \rangle }_{{\text{A, III}}}(s) &=&0,
\end{eqnarray}
\begin{eqnarray}
\rho^{\langle G^{3} \rangle }_{{\text{A, IV}}}(s) &=& \frac{\langle g_{s}^{3}G^3\rangle g_{s}^{2}}{2^{10}\pi^{4}} \int_{\alpha_{min}}^{\alpha_{max}} d\alpha\int_{\beta_{min}}^{\beta_{max}} d\beta\left\{ \frac{3F_{\alpha\beta}^{2}}{\alpha\beta}\right.
\nonumber\\
&-& \left.\frac{F_{\alpha\beta} (2s\alpha \beta+6m_{c}^{2})}{\alpha\beta}\right\},
\end{eqnarray}

\begin{eqnarray}
\rho^{\langle G^{3} \rangle }_{\text{A, V}}(s) &=& \frac{\langle g_{s}^{3}G^3\rangle g_{s}^{2}}{2^{7}\pi^{4}} \int_{\alpha_{min}}^{\alpha_{max}} d\alpha\int_{\beta_{min}}^{\beta_{max}} d\beta\left\{ -\frac{ m_{c}^{2}s(-\alpha-\beta+1)}{3}\right.
\nonumber\\
&+& \left.\frac{F_{\alpha\beta}(m_{c}^{2}(-2\alpha-2\beta+1) +6\alpha \beta s)}{12\alpha\beta}
-\frac{F_{\alpha\beta}^{2}}{8\alpha\beta}\right\}
\nonumber\\
&+&\frac{\langle g_{s}^{3}G^3\rangle g_{s}^{2}}{2^{10}\pi^{4}} \int_{\alpha_{min}}^{\alpha_{max}} d\alpha
\left\{ \frac{H_{\alpha}^{2}}{(-1+\alpha)\alpha}\right\},
\end{eqnarray}
\begin{eqnarray}
\rho^{\langle G^{4} \rangle }_{{\text{A, VI}}}(s) &=& \frac{\kappa \langle g_{s}^{2}G^{2} \rangle^{2}}{2^{9}\times 3\pi^2} \int_{\alpha_{min}}^{\alpha_{max}} d\alpha
\left\{ -(3H_{\alpha}-3m_{c}^{2}+(\alpha-1)\alpha s)\right\},
\end{eqnarray}
where we have used the following definitions:
\begin{eqnarray}
 F_{\alpha\beta}&=& m_{Q}^{2}(\alpha + \beta)-\alpha\beta s, \\
 H_{\alpha} &=& m_{Q}^{2}-s\alpha(1-\alpha), \\
 \alpha_{min} &=& \frac{1}{2} \left(1-\sqrt{1-\frac{4m_{Q}^{2}}{s}} \right), \\
 \alpha_{max} &=& \frac{1}{2} \left(1+\sqrt{1-\frac{4m_{Q}^{2}}{s}} \right),
\end{eqnarray}
and
\begin{eqnarray}
 \beta_{min} &=&\frac{m_{Q}^{2}\alpha}{-m_{Q}^{2} + s\alpha},\\
 \beta_{max} &=& 1-\alpha.
\end{eqnarray}

For case B, we have:
\begin{eqnarray}
\rho^{\text{pert}}_{{\text{B, I}}}(s) &=& \rho^{\text{pert}}_{{\text{A, I}}}(s), \\
\rho^{\langle G^{2} \rangle }_{{\text{B, II}}}(s) &=& \rho^{\langle G^{2} \rangle }_{{\text{A, II}}}(s), \\
\rho^{\langle G^{3} \rangle }_{{\text{B, III}}}(s)  &=& \rho^{\langle G^{3} \rangle }_{{\text{A, III}}}(s),
\end{eqnarray}

\begin{eqnarray}
\rho^{\langle G^{3} \rangle }_{{\text{B, IV}}}(s) &=& \frac{\langle g_{s}^{3}G^3\rangle g_{s}^{2}}{2^{13}\times 3\pi^{4}} \int_{\alpha_{min}}^{\alpha_{max}} d\alpha\int_{\beta_{min}}^{\beta_{max}} d\beta\left\{ \frac{F_{\alpha\beta}^{2}(33\alpha+33\beta -13)}{\alpha\beta}\right.
\nonumber\\
&+& \frac{F_{\alpha\beta} (-2s\alpha \beta(27\alpha +27\beta-20) + 2m_{c}^{2}(2\alpha + 2\beta -19))}{\alpha\beta}\nonumber\\
&+& \left.\frac{-8\alpha\beta m_{c}^{2}s(\alpha+\beta-1)+8\alpha^{2}\beta^{2}s^{2}(\alpha+\beta-1)}{\alpha \beta}\right\},
\end{eqnarray}

\begin{eqnarray}
\rho^{\langle G^{3} \rangle }_{{\text{B, V}}}(s) &=& \rho^{\langle G^{3} \rangle }_{{\text{A, V}}}(s), \\
\rho^{\langle G^{4} \rangle }_{{\text{B, VI}}}(s) &=& \rho^{\langle G^{4} \rangle }_{{\text{A, VI}}}(s).
\end{eqnarray}

For case C, we obtain:
\begin{eqnarray}
\rho^{\text{pert}}_{{\text{C, I}}}(s) &=& \frac{g_{s}^{4}}{ 2^{8}\times 9\pi^{6}} \int_{\alpha_{min}}^{\alpha_{max}} d\alpha\int_{\beta_{min}}^{\beta_{max}} d\beta \left\{ \frac{F_{\alpha\beta}^{3}m_{c}^{2}s(-\alpha-\beta+1)^{4}}{\alpha^{3}\beta^{3}} \right.\nonumber\\ &+& \frac{F_{\alpha\beta}^{5}(-\alpha-\beta+1)^{2}(-60(-\alpha - \beta+1)-45(\alpha+\beta))}{80\alpha^{4}\beta^{4}}\nonumber\\&-& \frac{F_{\alpha\beta}^{4}(-\alpha-\beta+1)^{2} (-30\alpha\beta s(-\alpha-\beta+1)+5m_{c}^{2}(\alpha + \beta)(-\alpha-\beta+1))}{16\alpha^{4}\beta^{4}}\nonumber\\
 &-& \left. \frac{F_{\alpha\beta}^{4}(-\alpha-\beta+1)^{2}(7m_{c}^{2}(-\alpha -\beta+1)^{2})}{16\alpha^{4} \beta^{4}}\right\},
\end{eqnarray}

\begin{eqnarray}
\rho^{\langle G^{2} \rangle }_{{\text{C, II}}}(s) &=& \frac{\langle g_{s}^{2} G^2 \rangle g_{s}^{2}}{2^{5}\times 3\pi^{4}} \int_{\alpha_{min}}^{\alpha_{max}} d\alpha\int_{\beta_{min}}^{\beta_{max}} d\beta  \left\{ \frac{-F_{\alpha\beta} m_{c}^{2}s(-\alpha -\beta+1)^{2}}{\alpha\beta}\right.
\nonumber\\
&+&\frac{F_{\alpha\beta}^{2}  (m_{c}^{2}(\alpha+\beta-1)(\alpha+\beta)-2\alpha\beta s(-\alpha-\beta+1))}{4\alpha^{2} \beta^{2}}\nonumber\\
&-& \left. \frac{F_{\alpha\beta}^{3}(4(-\alpha-\beta+1)+5(\alpha+\beta))}{12 \alpha^{2}\beta^{2}}\right\},
\end{eqnarray}

\begin{eqnarray}
\rho^{\langle G^{3} \rangle }_{{\text{C, III}}}(s) &=& \frac{\langle g_{s}^{3}G^3\rangle g_{s}^{2}}{2^{7}\times 3\pi^{4}} \int_{\alpha_{min}}^{\alpha_{max}} d\alpha\int_{\beta_{min}}^{\beta_{max}} d\beta\left\{\frac{ m_{c}^{2} s(-\alpha-\beta+1)^{2}(\alpha+\beta)}{\alpha\beta}\right.
\nonumber\\
&+& \frac{F_{\alpha\beta}(\alpha+\beta)(2m_{c}^{2} (-\alpha-\beta+1)(\alpha+\beta)+2\alpha \beta s(-\alpha-\beta+1))}{4\alpha^{2}\beta^{2}}\nonumber\\
&-& \left.\frac{F_{\alpha\beta}^{2}(\alpha + \beta)(-4(-\alpha-\beta+1)-5(\alpha +\beta))}{8\alpha\beta}
\right\},
\end{eqnarray}

\begin{eqnarray}
\rho^{\langle G^{3} \rangle }_{{\text{C, IV}}}(s) &=& \frac{\langle g_{s}^{3}G^3\rangle g_{s}^{2}}{2^{10}\times 3\pi^{4}} \int_{\alpha_{min}}^{\alpha_{max}} d\alpha\int_{\beta_{min}}^{\beta_{max}} d\beta\left\{ -\frac{11F_{\alpha\beta}^{2}(3\alpha+3\beta -2)}{\alpha\beta}\right.
\nonumber\\
&-& \frac{F_{\alpha\beta} (2s\alpha \beta(-27\alpha -27\beta+23) - 4m_{c}^{2}(\alpha + \beta -5))}{\alpha\beta}\nonumber\\
&-& \left.\frac{8\alpha\beta m_{c}^{2}s(\alpha+\beta-1)+8\alpha^{2}\beta^{2}s^{2}(\alpha+\beta-1)}{\alpha \beta}\right\},
\end{eqnarray}

\begin{eqnarray}
\rho^{\langle G^{3} \rangle }_{{\text{C, V}}}(s) &=& \frac{\langle g_{s}^{3}G^3\rangle g_{s}^{2}}{2^{10}\times 3\pi^{4}} \int_{\alpha_{min}}^{\alpha_{max}} d\alpha\int_{\beta_{min}}^{\beta_{max}} d\beta\left\{ -\frac{ 11F_{\alpha\beta}^{2}}{2\alpha\beta}\right.
\nonumber\\
&-& \left.\frac{F_{\alpha\beta}(9m_{c}^{2} -22\alpha\beta s)}{\alpha\beta}\right\}
\nonumber\\
&+&\frac{\langle g_{s}^{3}G^3\rangle g_{s}^{2}}{2^{11}\times 3\pi^{4}} \int_{\alpha_{min}}^{\alpha_{max}} d\alpha
\left\{ -35\frac{H_{\alpha}^{2}}{(1-\alpha)\alpha}\right\},
\end{eqnarray}

\begin{eqnarray}
\rho^{\langle G^{4} \rangle }_{{\text{C, VI}}}(s) &=& \frac{\kappa \langle g_{s}^{2}G^{2} \rangle^{2}}{2^{10}\times 3\pi^2} \int_{\alpha_{min}}^{\alpha_{max}} d\alpha
\left\{ 15H_{\alpha}+15m_{c}^{2}-7(1-\alpha)\alpha s \right\}.
\end{eqnarray}

\begin{eqnarray}
\rho^{\langle G^{5} \rangle }_{{\text{C, I}}}(s) =\frac{\langle g_{s}^{2}G^{2} \rangle\langle g_{s}^{3}G^{3}\rangle}{2^{8}\times \pi^2} \int_{\alpha_{min}}^{\alpha_{max}} d\alpha
\left\{ (3(1-\alpha)\alpha )\right\},
\end{eqnarray}
\begin{eqnarray}
\Pi^{\langle G^{5} \rangle }_{{\text{C, I}}}(M_{B}^{2}) = \frac{\langle g_{s}^{2}G^{2} \rangle\langle g_{s}^{3}G^{3}\rangle}{2^{8}\times 3\pi^{2}} \int_{0}^{1} d\alpha \frac{m_{c}^{2}(2m_{c}^{2}+5M_{B}^{2}\alpha(\alpha-1)) e^{-\frac{m_{Q}^{2}t_{\alpha}}{M_{B}^{2}}}}{\alpha(\alpha-1)M_{B}^{2}},
\end{eqnarray}
\begin{eqnarray}
\rho^{\langle G^{5} \rangle }_{{\text{C, II}}}(s) = - \frac{\langle g_{s}^{2}G^{2} \rangle\langle g_{s}^{3}G^{3}\rangle}{2^{10}\times \pi^2} \int_{\alpha_{min}}^{\alpha_{max}} d\alpha,
\end{eqnarray}
\begin{eqnarray}
\Pi^{\langle G^{5} \rangle }_{{\text{C, II}}}(M_{B}^{2}) = \frac{\langle g_{s}^{2}G^{2} \rangle\langle g_{s}^{3}G^{3}\rangle}{2^{9}\times 3\pi^{2}} \int_{0}^{1} d\alpha \frac{m_{c}^{2}e^{-\frac{m_{Q}^{2}t_{\alpha}}{M_{B}^{2}}}}{\alpha(1-\alpha)},
\end{eqnarray}
where $t_{\alpha}=\frac{1}{\alpha(1-\alpha)}$ and the terms $\Pi^{\langle G^{5} \rangle }_{{\text{C, I}}}(M_{B}^{2})$ and $\Pi^{\langle G^{5} \rangle }_{{\text{C, II}}}(M_{B}^{2})$ represent the contributions of the correlation funtion which have no imaginary parts but have nontrivial value under the Borel transform.

For case D, the results are:
\begin{eqnarray}
\rho^{\text{pert}}_{{\text{D, I}}}(s) &=& \rho^{\text{pert}}_{{\text{C, I}}}(s),\\
\rho^{\langle G^{2} \rangle }_{{\text{D, II}}}(s) &=& - \rho^{\langle G^{2} \rangle }_{{\text{C, II}}}(s),\\
\rho^{\langle G^{3} \rangle }_{{\text{D, III}}}(s) &=& - \rho^{\langle G^{3} \rangle }_{{\text{C, III}}}(s),
\end{eqnarray}
\begin{eqnarray}
\rho^{\langle G^{3} \rangle }_{{\text{D, IV}}}(s) &=& \frac{\langle g_{s}^{3}G^3\rangle g_{s}^{2}}{2^{13}\times 3\pi^{4}} \int_{\alpha_{min}}^{\alpha_{max}} d\alpha\int_{\beta_{min}}^{\beta_{max}} d\beta\frac{1}{\alpha\beta}\left\{F_{\alpha\beta}^{2}\left[ 96\alpha^{2} + 3\alpha(64\beta-55)\right.\right.\nonumber\\
 &+&\left. 96\beta^{2}-165\beta +53 \right] + 48s\alpha\beta(\alpha + \beta - 1)[ m_{c}^{2} + s\alpha\beta(7\alpha + 7\beta - 6)]\nonumber\\
 &-& 2F_{\alpha\beta}\left[ 3m_{c}^{2}(4\alpha + 4\beta + 1) + s\alpha\beta\left(384\alpha^{2} + \alpha(768\beta - 663) +384\beta^{2}\right.\right.\nonumber\\
  &-& \left.\left.\left.663\beta + 280\right)\right]
\right\},
\end{eqnarray}

\begin{eqnarray}
 \Pi^{\langle G^{3} \rangle }_{{\text{D, IV}}}(M_{B}^{2}) = \frac{\langle g_{s}^{3}G^3\rangle g_{s}^{2}}{2^{8}\times 3\pi^{4}} \int_{0}^{1} d\alpha \int_{0}^{1-\alpha} d\beta\frac{m_{c}^{6}(\alpha + \beta - 1)^2 (\alpha + \beta)^{3} e^{-\frac{f_{\alpha\beta}m_{Q}^{2}}{M_{B}^{2}}}}{\alpha^{2}\beta^{2}},
\end{eqnarray}

\begin{eqnarray}
\rho^{\langle G^{3} \rangle }_{{\text{D, V}}}(s) &=& \frac{\langle g_{s}^{3}G^3\rangle g_{s}^{2}}{2^{5}\times 3 \pi^{4}} \int_{\alpha_{min}}^{\alpha_{max}} d\alpha\int_{\beta_{min}}^{\beta_{max}} d\beta\left\{ -\frac{ 19F_{\alpha\beta}^{2}}{64\alpha\beta}\right.
\nonumber\\
&-& \left.\frac{F_{\alpha\beta}(m_{c}^{2}(\alpha + 17(1-\alpha - \beta) + \beta) -38\alpha\beta s)}{32\alpha\beta} + m_{c}^{2}s(1-\alpha-\beta)\right\}
\nonumber\\
&+&\frac{\langle g_{s}^{3}G^3\rangle g_{s}^{2}}{2^{11}\times 3\pi^{4}} \int_{\alpha_{min}}^{\alpha_{max}} d\alpha
\left\{ -5\frac{H_{\alpha}^{2}}{(-1+\alpha)\alpha}\right\},
\end{eqnarray}
\begin{eqnarray}
\rho^{\langle G^{4} \rangle }_{{\text{D, VI}}}(s) &=& \rho^{\langle G^{4} \rangle }_{{\text{C, VI}}}(s),
\end{eqnarray}
where $f_{\alpha \beta} = \frac{\alpha + \beta}{\alpha \beta} $ and the term $\Pi^{\langle G^{3} \rangle }_{{\text{D, IV}}}(M_{B}^{2})$ denotes the contribution of the correlation function which has no imaginary part but has nontrivial value under the Borel transform.

\end{widetext}

\end{document}